# Varying reference-point salience*

Alex Krumer, Felix Otto, & Tim Pawlowski

Alex Krumer
Faculty of Business Administration and Social Sciences,
Molde University College
e-mail: alex.krumer@himolde.no

Felix Otto (corresponding author)
Faculty of Economics and Social Sciences, Institute of Sports Science
University of Tübingen
e-mail: felix.otto@uni-tuebingen.de

Tim Pawlowski
Faculty of Economics and Social Sciences, Institute of Sports Science
University of Tübingen
e-mail: tim.pawlowski@uni-tuebingen.de

## Abstract

The salience of reference points may theoretically influence the loss aversion mechanism in effort provision. However, we still lack a direct test from real competitive settings that uses exogenous variation to measure the effect of salience. We exploit a natural experiment where highly professional and incentivized individuals perform in real competitive settings with exogenous variation of reference-point salience. While a relevant reference point is salient in some cases, it is obscured in others, which, under reasonable assumptions, may affect individuals' expectations. This enables us to examine the effect of reference-point salience on the loss aversion mechanism in effort provision. Our regression discontinuity analyses reveal that individuals with positive expectations outperform those with negative expectations, but only when the reference point is salient. (JEL D81, D84, D91, L83)

* We are grateful to Dietmar Fehr, Daniel Goller, Mario Lackner, Ignacio Palacios-Huerta, and Thomas Peeters for their helpful comments and discussions. We also thank seminar participants at the THE Workshop of the Tübingen-Hohenheim-Economics Association, the EEA Congress at the Bordeaux School of Economics, the Behavioral Workshop at Ulm University, the Workshop on Sports, Economics, and Natural Experiments at the University of Zurich, and the ESEA Conference on Sport Economics at University College Cork.

# 1 Introduction

Competition is a fundamental feature in economics. Many competitions are structured as contests and are often found in labor markets, R&D races, or political campaigns. Effort provision is essential in order to win in these contests but also very costly because "independently of success, all contestants bear some costs" (Moldovanu & Sela, 2001, p. 542). Thus, understanding effort provision in competitive environments is an important economic question.

Reference-dependent loss aversion is another fundamental feature in economic behavior, whose main idea is that losses are more painful than gains are enjoyable (Kahneman & Tversky, 1979; Tversky & Kahneman, 1992).[1] Thus, loss aversion has an important influence on effort provision in a way that individuals increase their effort more to avoid losses than to realize equally sized gains. As we can see, effort is a natural common element that connects reference-dependent loss aversion and contests.

For decades, much work has been done to identify the key factors that shape the formation of reference points.[2] In recent years, economists have been paying increased attention to salience of stimuli as a possible factor. While salience is a relatively new topic of research in economics, it has a long history in psychological research. Notably, already in 1982, Taylor and Thompson reviewed the accumulated evidence on attention in the psychological literature, describing salience as "the phenomenon that when one's attention is differentially directed to one portion on the environment rather than to others, the information contained in that portion

[1] See Brown et al. (2024) for a comprehensive meta-analysis on the loss aversion effect in a variety of settings.

[2] Relevant reference points in effort provision include personal bests/past peak performance (Anderson & Green, 2018; González-Díaz et al., 2024), round numbers (Allen et al., 2017), expectations (Abeler et al., 2011), or interim performance benchmarks (Berger & Pope, 2011; Pope & Schweitzer, 2011; Teeselink et al., 2023).

will receive disproportionate weighting in subsequent judgments" (Taylor & Thompson, 1982, p. 175).

According to a recent comprehensive review on salience in economic research by Bordalo et al. (2022), there are three main factors that may shape salience: (a) contrast, (b) surprise, and (c) prominence. We concentrate on prominence that "relates to the idea that stimuli highly available to our senses or in memory are more salient" (Bordalo et al., 2022, p. 524). As an example, Sigurdsson et al. (2009) manipulated the shelf placement of potato chips and found that the middle shelf was associated with the highest percentage of sales, suggesting that the middle shelf is the most salient.

Vast economic research on the role of salience in economic behavior only started appearing early in the 2010s. It has been shown that salience plays an important role in risky choice (Bordalo et al., 2012; Dertwinkel-Kalt & Köster, 2020; Vanunu et al., 2021), consumer choice (Bordalo et al., 2013, 2016, 2020; Dertwinkel-Kalt et al., 2017), finance (Cosemans & Frehen, 2021; Hartzmark & Solomon, 2019), and subjective performance evaluations (Jenter & Kanaan, 2015). There is also suggestive evidence that it may drive reference-point selection in the presence of several potential reference points (Markle et al., 2018; Pope and Schweitzer, 2011).[3] However, to the best of our knowledge, there are no studies that provide a direct test of the effect of salience on the loss aversion mechanism in effort provision in contests.

The aim of this paper is to fill the gap in understanding the effect of salience in contest-like settings, which is relevant for several strands in economic literature, including reference-dependent loss aversion, contest theory, and economics of attention.[4] The perfect way to investigate the effect of salience on the loss aversion mechanism in contests would be to have the exact same reference point, but to vary its salience à la the potato chips example mentioned above. However, it is extremely difficult to find a natural competitive environment where one can observe exogenous variation of reference-point salience and a clean identification of the loss aversion effect on either effort or performance.

[3] Note that Abeler et al. (2011) found that salience had no effect on effort provision in experimental non-contest-like settings.

[4] See Loewenstein and Wojtowicz (2025) for the comprehensive review on economics of attention.

In this paper, we use a unique opportunity to exploit an ideal natural experiment where we can observe exogenous variation of reference-point salience and a clean identification of the loss aversion effect on performance in a real competitive environment. Our setting involves highly professional individuals who perform in a contest with transparent and known rules and large monetary prizes. Moreover, a relevant reference point is salient in some cases, influencing individuals' performance expectations, while it is obscured in others.

More specifically, we take advantage of a change in rules that occurred in professional ski jumping. While this setting is rather unusual, it has the advantage of providing clean identification, as we will see in the following, and is therefore "the "perfect" domain" (List, 2020, p. 45).[5] In these competitions, 50 jumpers qualify for the main event based on their pre-event (qualification) ranks. During the main event, these athletes compete in the first round, from which the top 30 advance to the second (final) round. Only these 30 athletes compete in the second round and only they can win monetary prizes, making rank 30 a very prominent performance benchmark. While the qualification rankings are no longer relevant in the main event, we focus on pre-event (qualification) rank 30 as a reference point. This is because, as said above, rank 30 is a prominent benchmark that is likely to shape athletes' expectations about their current states of ability and their chances to advance to the final round. Thus, it is plausible to assume that athletes with a pre-event rank of 30 or better may have more favorable (i.e., positive) expectations of advancing to the final round, compared to athletes with a pre-event rank worse than 30.

[5] Using data from professional sports for economic research has many advantages. These include the fact that the participants compete under fixed and known rules with strong incentives to win and that the outcomes and the identities of the participants are fully observable (Bar-Eli et al., 2020; Palacios-Huerta, 2025).

Most importantly, what allows us to identify the variance in reference-point salience, is the following change in rules: Up to the 2016–17 season, the top 10 athletes were automatically pre-qualified and did not have to compete in the qualification round to be among the 50 jumpers in the main event. However, starting from the 2017–18 season, all athletes were required to compete in the qualification. This means that before the change, those who were nominally ranked 20 in the qualification were effectively ranked 30. After the change, those who were nominally ranked 30 in the qualification were also effectively ranked 30 (for more details, see Section 3). While the nominal qualification rankings should be irrelevant for athletes, previous studies have shown that even seemingly irrelevant information may play a significant role in decision-making (e.g., Bertrand et al., 2010; Kahneman et al., 1986). In our case, the difference between the nominal and effective rankings before the change could be sufficient for the irrelevant nominal rank 20 to receive more weight, increasing the perceived distance to the reference point.

It is therefore intuitive that identical nominal and effective values after the change make rank 30 more salient than before the change. This is in line with salience theory (Bordalo et al., 2012, 2013). On the one hand, athletes may overemphasize especially salient features (being nominally and effectively ranked 30 in the qualification, and thus putting more weight on the fact that their performance is very close to the prominent threshold of rank 30). On the other hand, they may underemphasize less salient but also important features (undervaluing their effective rank 30 because of being nominally ranked 20).

Our regression discontinuity analyses show that athletes with pre-event rank 30 or slightly above (e.g., 29) perform significantly better in the first round of the main event than those ranked slightly below (e.g., 31-32). This is in line with the reasoning that contestants with positive expectations increase their effort more to avoid losing their advantage compared to their close opponents. Importantly, however, we observe this performance discontinuity only in the period *after* the rule change when the nominal and the effective ranks were identical. We find no such effect in the period *before* the rule change when the difference between the nominal and effective pre-event ranks makes the reference point less salient. Therefore, we conclude that the cutoff at pre-event rank 30 acts as a reference point only when it is salient. In other words, the decreased salience makes the reference point less relevant for the formation of athletes' performance expectations – mitigating (or even eliminating) any incentive effect of loss aversion for athletes with pre-event ranks just above the elimination cutoff.

We contribute to the literature on reference-point salience by being the first to provide a direct test of how exogenous variation of reference-point salience affects the loss aversion mechanism in a real competitive environment. Our findings suggest that even highly professional individuals are affected by an incremental difference between nominal and effective values that obscures an otherwise salient reference point.

Our paper also contributes to the discussion about the loss aversion mechanism in contests. In this regard, previous studies provide conflicting empirical results. For example, Berger and Pope (2011) found that basketball teams who were narrowly losing at half-time had a significantly higher probability to win the match. However, Teeselink et al. (2023) could not replicate this finding in a variety of alternative settings.[6] Obviously, individuals who are slightly behind their goal (as reference point), and thus in the loss domain, should exert more effort than if the *same* individuals are slightly ahead. However, in contests between two *different* contestants (i.e., zero-sum games) it is theoretically not clear why the loss aversion mechanism would force a lagging contestant to exert more effort compared to the leading one. It is rather the other way around because the leading contestant has more to lose (the lead), and has, therefore, a greater incentive to increase effort. In contrast, the already lagging contestant can only gain and thus has weaker incentives.

[6] It is also worth mentioning Pope and Schweitzer (2011) who found that professional golf players performed significantly better when attempting for par (a typical number of shots that it takes to complete a hole) than when attempting for birdie (one shot less than par). However, the authors discuss the relevance of par as reference point in golf, which becomes less relevant towards the end of the contest and as a result loses half of its effect.

Following this line of reasoning, we add to the understanding of the role of expectation-based loss aversion in contests, which is driven by the idea that reference points may evolve endogenously from rational expectations, as presented in Kőszegi and Rabin (2006).[7] To illustrate, if we consider a zero-sum game with two symmetric contestants where one has an a-priori positive expectation regarding his win, whereas the other one's expectation is negative, the prediction would be that the contestant with positive expectations ends up with higher effort. This is because, similar to being in the lead, a contestant with positive expectations has more to lose than the opponent with negative expectations has to gain. This prediction is also theoretically supported by Fu et al. (2022), who show that in case of relatively equal contestants, expectation-based loss aversion reduces effort of a contestant with lower probability of winning whereas a contestant with higher probability of winning increases effort to avoid losing unexpectedly. One may also see an analogy to the endowment effect (Kahneman et al., 1990; Thaler, 1980) where the owner (a contestant in the lead) values the good (being in the lead) more than the one who considers purchasing it (the lagging contestant).[8] Our empirical analyses show that being slightly ahead (in expectation) enhances performance compared to being slightly behind (in expectations).

[7] While in general the empirical support for expectations-based reference points is mixed (e.g., Baillon et al., 2020; Fehr et al., 2022; Gneezy et al., 2017, Heffetz & List, 2014), several studies support their relevance for effort and performance (Abeler et al., 2011; Allen et al., 2017; Bartling et al., 2015; Gill & Prowse, 2012; Gill & Stone, 2010; Pope & Schweitzer, 2011).

[8] It is worth mentioning Hossain and List (2012) and Imas et al. (2017) who showed that upfront paid bonuses that can be lost in case of unsuccessful performance elicit higher efforts than bonuses that are paid after successful performance.

## 2 Stylized model

To guide our empirical approach, we first provide a stylized model of expectation-based loss aversion and reference-point salience to describe effort provision in contests. For that, we consider an all-pay contest structured as an ongoing zero-sum game with two players denoted by $i = 1, 2$. Player $i's$ value of winning the contest is $W_i$, which is common knowledge.[9] We assume symmetry between players, i.e., $W_1 = W_2 = W$. Each player exerts an effort of $x_i$. These efforts are submitted simultaneously, and the player with the higher effort wins the contest. Each player has a linear cost function $C(x_i) = x_i$.

We also assume that the players have different expectations of winning the contest, which evolve endogenously from rational expectations about future performance and the salience of the reference point. Such expectations typically relate to players' recent performances. Therefore, suppose the following:

Player 1 has positive expectations based on information about his recent performance, which was just above a certain reference point that would allow winning the contest. Player 2 has negative expectations based on his own recent performance that was just below the same reference point. Following the idea of loss aversion, according to which losing is more painful than winning is enjoyable, there are additional values of winning and losing as a function of previous expectations. Thus, if player 1, who has positive expectations about winning the contest, eventually loses, he will suffer a reduction of $d$ units from his payoff. However, the reference point must be salient so that the players may form their expectations in relation to this point. Thus, we introduce the salience parameter $s$ which is equal to one if the salience is strong enough, and zero otherwise. Accordingly, the reduction in utility due to a loss after having positive expectations can be presented as $d \cdot s$.

[9] All-pay structure is widely used to model contests. For example, Krumer et al. (2017) used the all-pay contest to study round-robin tournaments with three players. Their theoretical predictions on the first mover advantage were empirically confirmed by Krumer and Lechner (2017) who used data from Olympic wrestling tournaments. Furthermore, the empirical paper showed that in six out of seven possible cases, the all-pay model correctly predicted the identity of a wrestler with a higher probability of winning.

On the other hand, if player 2, who has negative expectations about winning the contest, eventually wins, he will gain additional $u$ units to his payoff. As previously, this happens if the salience of the reference point is strong. Therefore, the increase in utility due to a win after having negative expectations of winning can be presented as $u \cdot s$. In line with the loss aversion principle, we assume that $|d| > |u|$. In other words, a reduction in utility from losing with positive expectations is larger than an increase in utility from winning with negative expectations.

Taken together, if player 1 wins, his payoff is $W$. However, if he loses, his payoff is $-d \cdot s$. On the other hand, if player 2 wins, his payoff is $W + u \cdot s$. If he loses, his payoff is zero. Since $W + d \cdot s > W + u \cdot s$ and following Hillman and Riley (1989) and Baye et al. (1996), there is always a mixed-strategy equilibrium in which players randomize on the interval $[0, W + u \cdot s]$ to maximize their expected payoffs ($E_i$) according to their effort cumulative distribution functions $F_i, i = 1, 2$ that are implicitly given by:

$$E_1 = W \cdot F_2(x_1) - d \cdot s\big(1 - F_2(x_1)\big) - x_1 = (W + d \cdot s) \cdot F_2(x_1) - d \cdot s - x_1$$

$$E_2 = (W + u \cdot s) \cdot F_1(x_2) - x_2$$

In this mixed-strategy equilibrium, players' expected efforts are equal to:

$$x_1 = \frac{(W + d \cdot s)^2 \cdot (W + u \cdot s)}{(2W + d \cdot s + u \cdot s)}$$

$$x_2 = \frac{(W + u \cdot s)^2 \cdot (W + d \cdot s)}{(2W + d \cdot s + u \cdot s)}$$

And, since $d > u$, player 1's probability of winning is given by:

$$p_1 = 1 - \frac{W + u \cdot s}{2 \cdot (W + d \cdot s)}$$

whereas player 2's probability of winning is:

$$p_2 = \frac{W + u \cdot s}{2 \cdot (W + d \cdot s)}$$

The following proposition summarizes the effects of the salience of the reference point and expectation-based loss aversion. Obviously, in the absence of a salient reference point (*s=0*), the difference in players' expectation-based loss aversion is not relevant for the players' effort provision and their probabilities of winning.

**Proposition** *If the reference point is salient ($s = 1$), the difference in players' expectation-based loss aversion, where $d > u$, affects players' efforts and their probabilities of winning such that the player with a positive expectation (player 1) exerts a higher effort and, thus, has a higher probability of winning ($p_1 > 0.5$).*

To further illustrate this proposition in the Prospect Theory framework (Kahneman & Tversky, 1979), Figure 1 provides a visualization of the value functions with positive and negative expectations (for simplicity, without diminishing sensitivity). In both graphs, we see that the line is steeper in the loss domain than in the gain domain. In the graph on the left, we see that a player with positive expectations suffers from a loss more than without expectations, as illustrated by the steeper value function (dotted line). There is no such difference in the gain domain, where the value functions overlap. In the graph on the right, we see that a player with negative expectations values a gain more than without expectations, while there is no difference in the loss domain. Lastly, comparing the value functions from both graphs, we see that a player with positive expectations experiences a larger difference in valuation between the gain and the loss domains than a player with negative expectations. This shows that a player with positive expectations has higher incentives to exert effort to avoid a loss than a player with negative expectations to achieve a win.

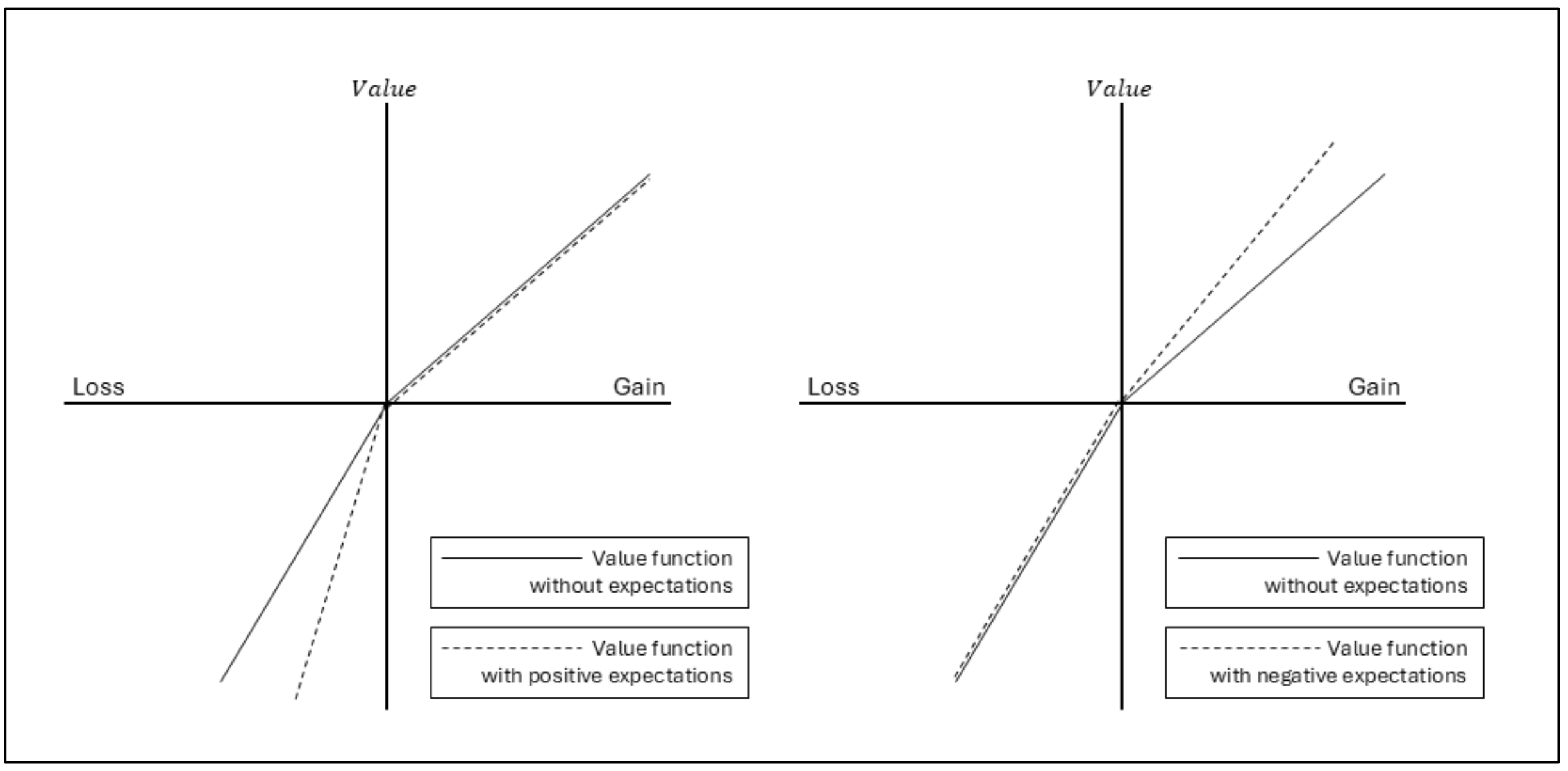

**Figure 1.** Loss aversion with and without positive and negative expectations.

# 3 Description of ski jumping World Cup competitions

## 3.1 Contest design

In ski jumping, athletes perform jumps from a ski slope, which is on a hill. The slope consists of a steep track to generate speed and a take-off ramp. The hill is used for landing the jumps. The athletes' aim is to maximize their jumping distance and style points. Both performance determinants are mutually dependent because ski jumping is a highly technical discipline, which requires strength, speed, and coordination to perform complex movement sequences in a short moment of time. Performances are therefore highly standardized and produced through the interplay between physical effort (strength and coordination) and cognitive effort (concentration and focus), while external factors like wind conditions are taken into account.

Jumping distance is measured in meters with intervals of 0.5 and converted to a jumping distance point score that relates to the hill size. To calculate the distance points, each hill has a predetermined construction point, called the K-point, in the landing area. At the most common competitions, athletes receive 60 points for landing on the K-point and 1.8 or 2.0 points are added (deducted) for each additional meter beyond (below) the K-point at large or normal hills, respectively. Style points are based on predefined judging criteria and are awarded by a judging panel, which consists of five judges, each of whom can award between zero and 20 points for a jump with intervals of 0.5 points. The lowest and highest scores are truncated, and the remaining three scores are summed up, yielding a maximum style-point score of 60 for a perfectly styled jump. To increase safety and fairness, wind and gate points are added to the total score of each jump. Wind points capture (dis-)advantages in wind conditions, whereas gate points capture changes in the starting gate of the ski slope, and thus (dis-)advantages in generating speed. These components also allow us to capture external factors beyond ability and effort that may affect individual performance.

The international governing body, Fédération Internationale de Ski (FIS), has organized international World Cup (WC) seasons for men during each Northern Hemisphere winter period since 1979. Ordinary WC competitions have an all-against-all contest design that consists of a qualification round in the pre-event phase and two rounds in the main event phase, as shown in Figure 2 for both periods before and after the rule change.[10] Athletes always perform one jump per round. The top 50 athletes in the qualification round advance to the main event and compete in Round 1 (as illustrated by the dashed lines below rank 50 in Figure 2). The top 30 athletes from Round 1 then proceed to Round 2 and compete for the win (as illustrated by the dashed lines below rank 30 in Figure 2). In this regard, the point scores and rankings in qualification are no longer relevant for competition in the main event, whereas the total point scores from Rounds 1 and 2 are equally important in determining the final ranks of the top 30 athletes in the main event. This is an important feature of the ski jumping setting because, in principle, all athletes have equally strong incentives to provide maximum effort in both rounds, which eliminates any strategic component or loafing in effort provision between rounds. Moreover, the top 30 athletes receive prize money[11] and WC points, which are added to the season's WC standings.

[10] Non-ordinary World Cups are flying hill competitions and the Four Hills Tournament, both of which have a different contest design and are thus excluded from our study.

| Before rule change | | | | After rule change | | | |
|---|---|---|---|---|---|---|---|
| *Pre-event phase* | | *Main event phase* | | *Pre-event phase* | | *Main event phase* | |
| Quali rank | Pre-event rank | Event round 1 rank | Event round 2 final rank | Quali rank | Pre-event rank | Event round 1 rank | Event round 2 final rank |
| *Top 10 in World Cup standings* | 1 ⋮ 10 | 1 ⋮ 10 | 1 ⋮ 10 | 1 ⋮ 10 | 1 ⋮ 10 | 1 ⋮ 10 | 1 ⋮ 10 |
| 1 ⋮ 20 | 11 ⋮ 30 | 11 ⋮ 30 | 11 ⋮ 30 | 11 ⋮ 30 | 11 ⋮ 30 | 11 ⋮ 30 | 11 ⋮ 30 |
| 21 ⋮ 40 | 31 ⋮ 50 | 31 ⋮ 50 | *keep ranks from round 1* | 31 ⋮ 50 | 31 ⋮ 50 | 31 ⋮ 50 | *keep ranks from round 1* |
| 41 and worse | >=51 | *eliminated from main event* | | >=51 | >=51 | *eliminated from main event* | |

**Figure 2.** Description of the ski jumping World Cups' contest design and the qualification procedures before and after the rule change.

Notes. Elimination cutoffs after qualification and first round are marked with dashed lines.

### 3.2 Relevance of setting: Reference point and rule change

In this setting, being ranked 30 or better is practically irrelevant in the qualification because all top 50 athletes qualify for the main event.[12] However, it becomes very important in Round

[11] In our sample period, prize money ranges from CHF 100 for rank 30 to CHF 10,000 for the WC event winner.

[12] It is also irrelevant for the starting order in Round 1, which is determined by the reversed order of WC standings for athletes with a WC rank and by draw for athletes without points (same procedure is used for qualifications).

1 because, as described above, only the top 30 advance to Round 2. These top 30 athletes get the chance to win the event, receive prize money, and improve in the overall WC standings. Thus, rank 30 is a very well-known and prominent threshold in ski jumping.

Although the qualification results no longer matter for the competition in the main event, it is plausible to assume that pre-event performances raise strong expectations about subsequent performances among athletes. This would imply that athletes with a pre-event rank of 30 or better expect to perform similarly in Round 1 and consequently have positive expectations to make it to Round 2. In this regard, however, those athletes who are ranked exactly 30th in the qualification or just slightly better must fear elimination after Round 1 because athletes with a pre-event rank of 31 or slightly worse are only slightly behind in terms of their previous performances. In such a competitive setting, we assume that athletes are sensitive to deviations from their expectations, where losses relative to expectations loom larger than equally sized gains, as emphasized by Kőszegi and Rabin (2006) and described in our stylized model ($|d| > |u|$). Hence, we focus on this elimination cutoff *after* Round 1 as an expectation-based and salient reference point for athletes' effort provision and performance *before* they actually compete in this round (as illustrated by the wavy lines below rank 30 in Figure 2).

Besides the potential relevance of performance expectations and pre-event rank 30 as a reference point, there was a rule change in the qualification procedure before the 2017/18 season that changed the nominal values of ranking information in the qualification phase (as illustrated by the quali ranks-columns before and after the rule change in Figure 2). Before the change, the 10 highest ranked athletes in the ongoing WC standings were automatically prequalified for the main event. These top athletes did not compete in the qualification and could instead jump for training purposes or could not jump at all.[13] The new rule from the 2017/18 season onwards requires all athletes to compete in the qualification.

As described in Figure 2, this means that before the rule change, an athlete who was nominally ranked 20 in the qualification round was effectively ranked 30 in the underlying pre-event performance ranking. Importantly, the nominal and effective values of pre-event ranks are well known to all athletes in both periods. As such, any difference between them is easy to understand and obvious to all athletes when forming performance expectations. However, the difference between nominal qualification ranks and effective pre-event ranks may increase the *perceived* distance to the cutoff so that it appears less relevant (i.e., being further from the elimination cutoff) if an athlete puts more weight to the nominal ranking. As such, the difference in nominal and effective values of pre-event ranks may obscure the perceived salience of the reference point and thus the incentive effect of loss aversion in effort provision. After the rule change, the qualification ranks align with the pre-event ranks, such that the nominal ranks correspond to the effective ranks. This means that athletes know in both situations whether they are relative to the cutoff point of being eliminated after Round 1, but this information becomes more salient after the rule change.

# 4 Data and variables

## 4.1 Data collection and sample

We collected data from the World Cup seasons 2015 to 2020, including three seasons before and three seasons after the rule change to generate a balanced dataset.[14] The last 2020 season ended slightly earlier due to the COVID-19 pandemic. The data were retrieved from the official result protocols provided on the FIS website. We only use data from ordinary WC competitions on normal and large hills because their contest design allows for performance comparisons. We further restrict the data by only including events where a qualification round and both rounds in the main event were held. We also exclude athletes who were disqualified, did not start, or did not finish in a given event.

[13] Prequalified athletes who jumped did also not receive total point scores, with one exception of two events in the 2017/18 season that where part of a competition series (titled Raw Air), where the qualification counted for the overall classification of the series. Results remain if we remove these events from the sample.

[14] For simplicity, we refer to the latter year to label each season (i.e., the 2014–2015 WC season is labeled as the 2015 season).

As summarized in Table 1, the sample period before the rule change covers 53 WC events where 164 different athletes performed 2,629 jumps in Round 1. The sample period after the rule change covers 44 WC events where 140 athletes performed 2,161 jumps in Round 1. The lower number of events after the rule change is due to a reduced schedule in the WC season in 2018 because of the Winter Olympics in Pyeongchang and the beginning of the COVID-19 pandemic crisis in March 2020. Overall, our dataset includes a total of 4,790 performance observations in Round 1 of WC events.

**Table 1.** Sample size.

| | Before rule change (2015–2017 seasons) | After rule change (2018–2020 seasons) |
|---|---|---|
| Number of World Cups | 53 | 44 |
| Number of athletes | 164 | 140 |
| Number of jumps | 2,629 | 2,161 |
| Total no. of obs. (jumps) | 4,790 | |

### 4.2 Description of variables

To analyze ski jumping performances as a function of pre-event ranks, we measure performance by using a dummy variable that is equal to one if an athlete advances to Round 2 and zero otherwise.[15] For reasons of consistency across periods, we will use the effective pre-event ranks and not the nominal qualification ranks as a variable to measure the differences in pre-event ranks. To measure athletes' abilities, we use the official WC standing points before a competition (Harb-Wu & Krumer, 2019).[16] We also use the previous event rank of athletes achieved in their preceding competition to additionally capture their current form.[17] To capture home advantage that plays a significant role in ski jumping (Krumer et al., 2022), we consider whether athletes compete at an event in their home country.

[15] We also consider the absolute performance measures – that is, jumping distance in points and style points – in further analyses (see Appendix C).

Table 2 provides an overview and summary statistics of variables for both subsamples. In line with the ski jumping rules, about 60 percent of athletes (top 30 out of 50 participants) advance from Round 1 to Round 2. Note that the average WC standing points are lower after the rule change due to the lower number of events. The mean values of the other variables are comparably similar in both subsamples.

**Table 2.** Descriptive statistics of the variables.

| | Before rule change (2015–2017 seasons) | | After rule change (2018–2020 seasons) | |
|---|---|---|---|---|
| Variable | Mean (SD) | Min-Max | Mean (SD) | Min-Max |
| Advance to Round 2 (yes = 1) | 0.60 | 0–1 | 0.61 | 0–1 |
| Pre-event rank | 25.47 (14.41) | 1–50 | 25.38 (14.42) | 1–50 |
| WC standing points | 181.86 (285.26) | 0–2303 | 168.37 (263.32) | 0–2085 |
| Previous event rank[1] | 24.17 (14.08) | 1–51 | 23.95 (13.97) | 1–51 |
| Home event (yes = 1) | 0.12 | 0–1 | 0.11 | 0–1 |
| No. of obs. | 2,629 | | 2,161 | |

Notes. Standard deviations (SD) are reported in parentheses for metric variables. [1] This variable only includes 2,302 and 1,865 valid observations for the samples before and after the rule change, respectively.

[16] For the first competition of each season, we use the WC points from the final WC standings from the previous season.

[17] This covariate includes missing values for the first competition of each season. Missing values for single cases can also occur if athletes did not compete in the main event of a previous World Cup. This reduces the number of observations when using this variable in data-driven RD window selection that will be presented in the next sections.

# 5 Empirical strategy

## 5.1 Comparison of performances before and after the rule change

We start our analysis by comparing athletes' performances before and after the rule change to examine whether and how the change in nominal values of pre-event ranks is associated with performance. As shown in Figure 3, we visually compare the probability of advancing to Round 2 (on the y-axis) as a function of pre-event ranks (on the x-axis) between the two sample periods. For a better overview, we combined the pre-event ranks into groups of five. As expected, we see a fairly linear decrease in performance if we go down the ranking groups in the period *before* the rule change. We observe a similar pattern *after* the rule change; however, there is a jump in performance for those ranked 26–30 compared to the sample period before the rule change. The difference in performance between the two periods for these ranks is statistically significant (mean difference = -0.115, *p*-value = 0.010; see also Table A1 in Appendix A) and evident in absolute terms as expressed in total point scores (see Figure A1 and Table A2 in Appendix A). These findings serve as a first indication that the cutoff between pre-event ranks 30 and 31 might be a relevant reference point *after* the rule change, when loss aversion might affect performances of athletes ranked 30 or slightly better.

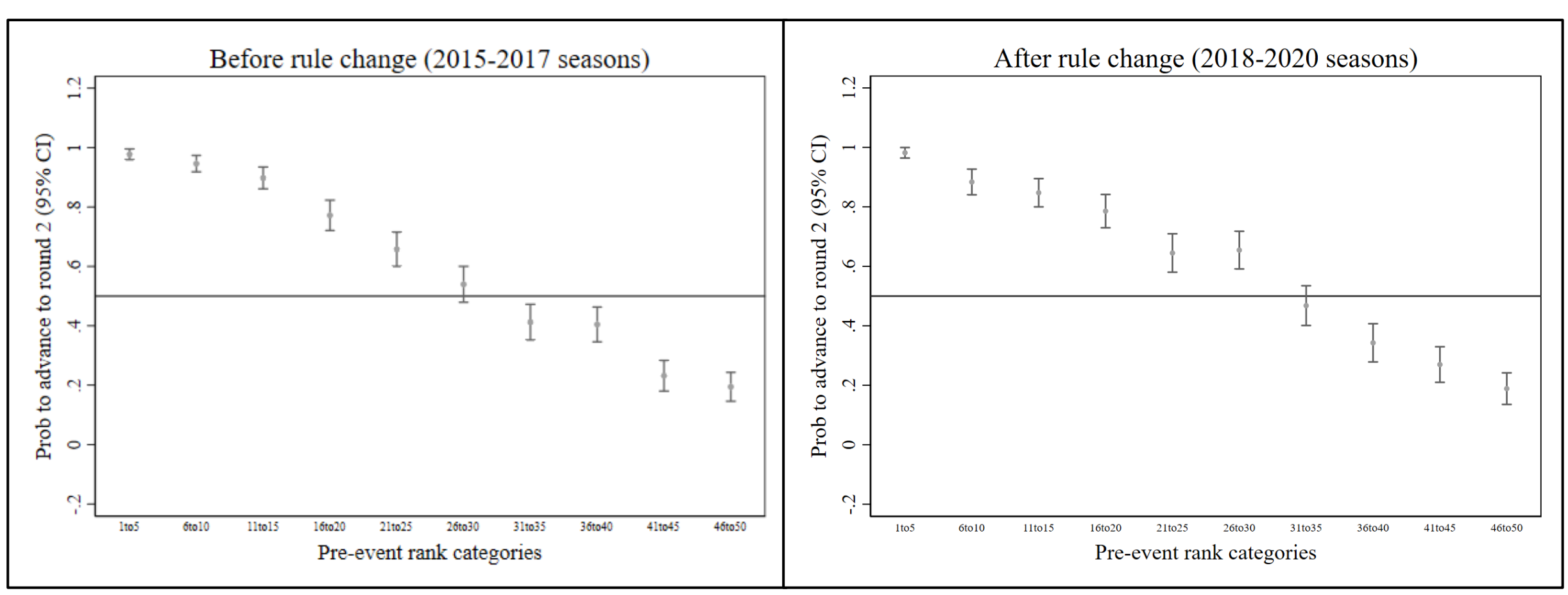

**Figure 3.** Comparison of performances in Round 1 as a function of ski jumpers' pre-event ranks.

Alternatively, one could also think that athletes who competed in the qualification *before* the rule change might have neglected the existence of the prequalified top 10 athletes and only considered their nominal ranks. In this case, athletes with an effective pre-event rank of 36–40 but a nominal qualification rank of 26–30 would use the cutoff point as their reference point. In Figure 3, we actually observe a jump in performance of this rank group in the period before the rule change in comparison to ranks 41-45. However, when comparing ranks 36-40 before and after the change we find no statistically significant difference at conventional levels (mean difference = 0.062, $p$-value = 0.165; see also Table A1 in Appendix A).

A general concern with this simple comparison is that we do not compare the same type of athletes in terms of ability before and after the rule change. This is because some top athletes who were automatically prequalified before the change may exert less effort in the qualification after the change than athletes outside the top 10, since they will most likely qualify anyway. Therefore, they might receive worse pre-event ranks when they compete in the qualification. In fact, while an athlete from the top 10 is naturally ranked on average fifth before the rule change (because he was automatically prequalified), we find that after the rule change, an athlete from the top 10 is ranked on average only $12^{th}$ in the qualifying round. We further try to examine this potential selection issue by comparing the WC standing points of the pre-event rank groups between the two sample periods (see Table A3 in Appendix A). As expected, selection predominantly exists in the top rank groups, whereas the WC standing points of the relevant rank groups (that is, pre-event ranks 26–30 and 31–35) are not statistically different. We also find that 33 athletes (20% of athletes) make 50% of jumps within the group of pre-event ranks 26-35 before the rule change, and these same athletes still make 42% of jumps of that pre-event rank group after the change. Still, however, this simple comparative approach does not make it possible to distinguish between selection effects and reference point effects.

To overcome this issue, our identification strategy is based on the idea of not just comparing pre-event ranking effects *between* the two sample periods, but comparing athletes ranked just below the reference point to those ranked just above *within* the same period. Intuitively, this within-comparison is plausible because it is reasonable to assume that athletes with pre-event ranks around the cutoff are very similar in ability and thus comparable within each of the two periods.[18] Moreover, athletes cannot *strategically* influence their pre-event rank because it depends on the relative performances of all the other athletes in the qualification round. Knowing all other performances, including many athletes who still need to jump, and then meeting the exact requirements to get a particular rank is virtually impossible. We can therefore assume quasi-random assignment around the cutoff and compare performances of athletes with pre-event ranks close to rank 30 within each period. This allows us to identify whether this cutoff rank acts as a reference point separately in each period by employing a regression discontinuity (RD) approach.

### 5.2 Regression discontinuity approach

The RD approach is a credible method for causal inference in situations where a treatment is quasi-randomly assigned based on a cutoff that relates to a score (Lee & Lemieux, 2010). As discussed above, assuming quasi-random assignment to the treatment condition is plausible in our setting, specifically for those with a pre-event rank close to the elimination cutoff. This allows us to examine the discontinuous change in performances between athletes who are ranked at or just below and above this cutoff.

In our setting, the score consists of the athletes' pre-event ranks and denotes the running variable. The athletes' reference point denotes the RD cutoff, which is between pre-event ranks 30 and 31. As discussed above, we assume that athletes with pre-event rank 30 or better have more positive expectations regarding their advancement to the second round than athletes with pre-event rank 31 or worse. Thus, following the idea that $|d| > |u|$, as presented in the stylized model, the pain of athletes with positive expectations who fail to advance to the second round would be higher than the joy of athletes with negative expectations who advance. In other words, athletes ranked 30 or better belong to the treatment group that is expected to experience loss aversion when performing in Round 1 of the main event, compared to those ranked 31 or worse (control group). We therefore consider the following baseline RD model:

$$Y_{ie} = a + \tau \mathbf{1}(rank_{ie} < c) + f(rank_{ie}) + \epsilon_{ie},$$

where $Y_{ie}$ is the performance outcome of athlete $i$ in Round 1 of event $e$, $\mathbf{1}(rank_{ic} < c)$ is an indicator function that takes the value of one for athletes who are below the cutoff $c$ (i.e., pre-event ranks ≤ 30) and thus treated, and zero for athletes who are above $c$ (i.e., pre-event ranks ≥ 31). Here, $f(rank_{ie})$ is a discrete function of pre-event ranks on each side of $c$. We estimate the treatment effect $\tau$ separately for both sample periods. In addition, we employ a difference-in-discontinuities (Diff-in-Disc) approach to estimate the difference in the discontinuity at $c$ before and after the rule change.

[18] We also support this within-comparison approach empirically in Appendix B.

In our main analysis, we employ local randomization RD, which is used if the running variable ($rank_{ie}$) is a discrete score and includes relatively few distinct mass points to identify treatment effects (Cattaneo & Titiunik, 2022). The local randomization framework requires a well-defined window $W = [c - \omega, c + \omega]$ around cutoff $c$ in which subjects are 'as good as' randomly assigned by having the same probability of receiving one of the scores. This is most likely to hold in the smallest possible window $W = [c - 1, c + 1]$, which includes pre-event ranks 30 and 31. While considering larger windows is usually not necessary, exploiting a larger number of effective observations to compare treated and control groups can show the sensitivity of results regarding the window choice.[19] Besides, the incentive effect of loss aversion should also exist for athletes that are ranked slightly better than pre-event rank 30 as they may also fear elimination after Round 1. Therefore, in alternative specifications, we consider the next larger window [29, 32] and employ a data-driven window selection procedure to consider the largest possible window in which athletes are comparable. The objective and transparent data-driven window selection uses our performance-related variables as predetermined covariates to implement covariate balance tests, suggesting a window $W$ in which athletes do not systematically differ in terms of ability.[20] $W$ is the window furthest away from the cutoff where the minimum $p$-values of the balance tests are >=0.15 for this and all nested windows (Cattaneo et al., 2023a).

In addition, we estimate the treatment effect with the continuity-based RD approach, which is a common alternative to complement local randomization RD estimates if the number of mass points in the discrete running variable is moderate. We also use this approach to provide a formal statistical test of the difference in the treatment effect before and after the rule change in a Diff-in-Disc design, as first proposed by Grembi et al. (2016). We employ non-parametric local polynomial approximation, where the treatment effect is estimated by fitting local linear regressions for observations inside a bandwidth where observations closer to the cutoff receive more weight than those further away. For estimation and inference, we follow Cattaneo et al. (2020), using a triangular kernel function to assign weights, data-driven common mean squared error (MSE) optimal bandwidth selection with the resulting MSE-optimal point estimator, and robust bias-corrected inference.

[19] Moreover, athletes with the same number of total points in the qualification also share the same pre-event rank, which may slightly distort the rank distribution.

[20] We also used alternative ability measures, such as the WC standing ranks, WC standing points standardized by events, and a measure that considers the final event ranks from the last five competitions. They all generated similar windows and results. These results are available upon request.

Lastly, our RD approach can be further supported by validation and falsification tests, including treatment effect tests on the predetermined covariates, assessing the density of the running variable, and estimations at placebo cutoffs (Cattaneo & Titiunik, 2022; Cattaneo et al., 2023a). We provide and discuss these tests in Appendix B.

# 6 Results

We start by visualizing the RD design and results in RD plots. In Figure 4, the upper plots show the global polynomial fit, represented by a solid line, and the local sample mean of each pre-event rank, represented by dots. The vertical lines mark cutoff $c$ and the vertical distances between the two conditional expectations at $c$ indicate the treatment effects under continuity conditions. In the lower plots in Figure 4, the solid lines show the local polynomial fit of degree zero; that is, the constant within the largest suggested windows from the data-driven selection procedure (windows [28, 33] and [26, 35] for the samples before and after the rule change, respectively). The constant vertical distance between the constants below and above $c$ indicate the treatment effect under local randomization conditions.

Without going into the question of statistical difference (that will be done later), all RD plots show that the probability to advance to Round 2 changes discontinuously at the cutoff, being higher for treated athletes just below the cutoff in the samples before and after the rule change. However, the vertical distance at the cutoff is considerably larger after the rule change under both continuity as well as local randomization conditions.

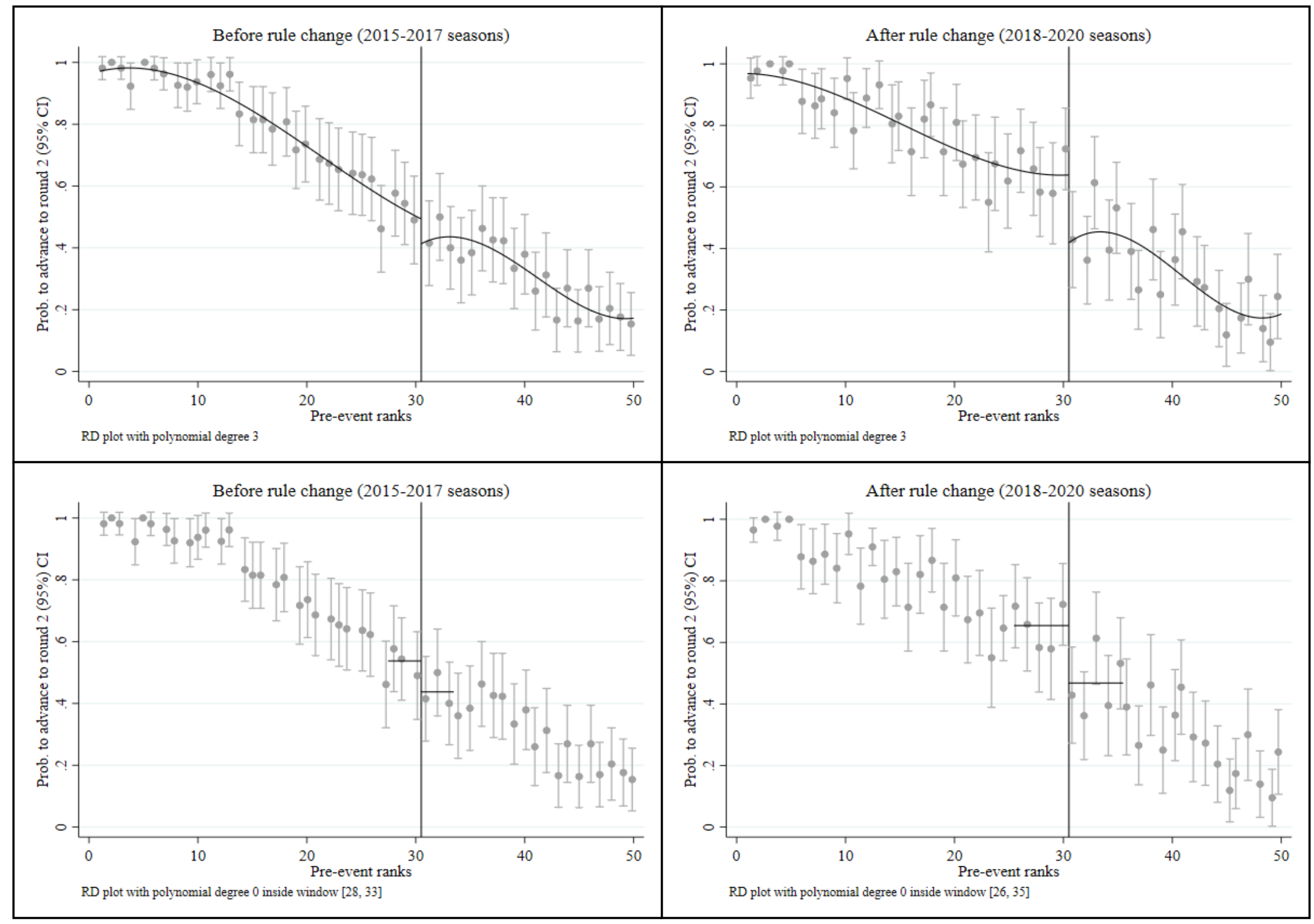

**Figure 4.** RD plots.

Notes. The dots mark the sample means within bins (that is, pre-event ranks) with 95% confidence intervals. Vertical lines mark the cutoff between rank 30 and 31 ($c = 30.5$). In the upper plots, the lines display the polynomial fit of degree 3 on each side of the cutoff. In the lower plots, the lines display the polynomial fit of degree 0 (that is, as a constant), within the relevant windows on each side of the cutoff.

## 6.1 Local randomization RD

We describe the treatment effect more formally in Table 3, using local randomization analysis with the three different windows, as described in Section 4.2. For the sample period before the rule change, we find that the probability to advance to Round 2 increases, on average, by 7.5 percentage points for treated athletes with pre-event rank 30 compared to those with pre-event rank 31 (see Column 1). However, this effect is statistically insignificant, suggesting that the cutoff has no impact on performance when the nominal and effective pre-event ranks differ. The effect remains insignificant and similar in size when we consider specifications with larger windows (Columns 2 and 3). This suggests that the cutoff does not seem to be a salient reference point that considerably affects athletes' performances via the loss aversion mechanism.

**Table 3.** Local randomization RD estimates on advancing to Round 2.

| | Before rule change (2015–2017 seasons) | | | After rule change (2018–2020 seasons) | | |
|---|---|---|---|---|---|---|
| | (1) | (2) | (3) | (4) | (5) | (6) |
| Point estimate | 0.075 | 0.061 | 0.100 | 0.295 | 0.266 | 0.187 |
| P-value | 0.516 | 0.476 | 0.102 | 0.008 | 0.000 | 0.000 |
| Window | [30, 31] | [29, 32] | [28, 33] | [30, 31] | [29, 32] | [26, 35] |
| Effective no. of obs. (treated / controls) | 51 / 53 | 108 / 105 | 160 / 160 | 47 / 42 | 85 / 89 | 220 / 218 |
| No. of obs. | | 2,629 | | | 2,161 | |

Notes. The dependent variable is a dummy denoting if a ski jumper advances to Round 2 in the main event. The running variable is the pre-event rank and the cutoff is between rank 30 and 31. The windows for RD analyses in Columns 3 and 6 derived from optimal window selection based on the predetermined covariates WC standing points, previous event rank, and home event. Point estimates report the difference in means and *p*-values derived from Fisherian simulation-based methods.

The pattern is much more distinct in the sample period after the rule change. In Column 4, we find that the probability to advance to Round 2 increases, on average, by approximately 30 percentage points for athletes with pre-event rank 30 compared to those with pre-event rank 31. The effect size becomes smaller when we consider larger windows (Columns 5 and 6), which is in line with the diminishing sensitivity component of Prospect Theory (Kahneman & Tversky, 1979). Despite a decrease in the size of the effect, the estimates remain statistically significant and substantial in size. Thus, the cutoff seems to be a prominent reference point such that the relative performance of athletes with pre-event ranks of 30 or slightly better is superior to that of athletes with pre-event ranks just below the cutoff.

### 6.2 Continuity-based RD and difference-in-discontinuities approach

To complement the previous analysis, we also estimate the treatment effect with the more common continuity-based RD approach. In Table 4, we report the MSE optimal point estimates and robust p-values (Cattaneo et al., 2020).[21] The effects are estimated within the MSE-optimal bandwidth that includes pre-event ranks 25 to 36 and are thus local but also cover athletes slightly further away from the reference point. As previously, before the rule change, we find no significant difference in performance between athletes with pre-event ranks slightly above and below the cutoff (Column 1). After the rule change, however, we find that athletes with better pre-event ranks had a 27 percentage points higher probability to advance to Round 2 than athletes with worse pre-event ranks. The effect is statistically significant at the 1% level (Column 4).

[21] P-values from conventional inference are similar to the robust bias-corrected method and can be found in the supplementary replication material.

**Table 4.** Continuity-based RD estimates on advancing to Round 2.

| | Before rule change (2015–2017 seasons) | | | After rule change (2018–2020 seasons) | | |
|---|---|---|---|---|---|---|
| | (1) | (2) | (3) | (4) | (5) | (6) |
| Point estimate | 0.046 | 0.047 | 0.043 | 0.274 | 0.302 | 0.288 |
| | (0.089) | (0.077) | (0.080) | (0.090) | (0.089) | (0.010) |
| P-value | 0.669 | 0.556 | 0.650 | 0.003 | 0.001 | 0.005 |
| Bandwidth | [25, 36] | [24, 37] | [22, 39] | [25, 36] | [25, 36] | [26, 35] |
| Effective no. of obs. (treated / controls) | 320 / 316 | 373 / 370 | 422 / 404 | 262 / 259 | 262 / 259 | 188 / 184 |
| No. of obs. | | 2,629 | | | 2,161 | |

Notes. The dependent variable is a dummy denoting if a ski jumper advances to Round 2 in the main event. The running variable is the pre-event rank, and the cutoff is between ranks 30 and 31. The continuity-based RD analyses estimate local (first-order) polynomial regressions with a triangular kernel function to assign weights to the observations and common mean squared error (MSE)-optimal bandwidth selection, reporting the MSE-optimal point estimates. Standard errors are clustered at the athlete level and presented in parentheses. Reported *p*-values are obtained from robust bias-corrected inference. The RD analyses in Columns 2 and 5 use covariate adjustment based on the predetermined covariates WC standing points and home event; the RD analyses in Columns 3 and 6 additionally include previous event rank as covariate.

To further increase efficiency and precision of statistical inference, we also estimate local linear RD with covariate adjustment (Cattaneo et al., 2023b). First, we present the results after including the predetermined covariates, which are WC standing points and home event (Columns 2 and 5 in Table 4). We then add the previous event rank as an additional covariate. Note that there are missing values for the first events of each season, which slightly reduces the number of observations (Columns 3 and 6 in Table 4). The point estimates remain similar in magnitude and precision in the samples before and after the rule change. Overall, the results in Table 4 are in line with the local randomization RD analysis (Table 3). In both cases, we find that athletes only perform significantly better if they are ranked better than the cutoff *after* the rule change.

In addition to our main RD analysis, we also estimate a Diff-in-Disc specification to compare treatment effects before and after the rule change. For that, we pool the samples from the two periods and estimate a local linear RD with triangular kernel and MSE-optimal bandwidth selection. We include an interaction term with an indicator function that denotes the periods before and after the rule change. In Table 5, we report the difference in discontinuities, that is the estimated increase in the treatment effect from the period before the rule change to the period after the change. As before, we estimate specifications without including predetermined covariates (Colum 1 in Table 5) and with covariate controls (Column 2 and 3 in Table 5). The interaction coefficients show that the probability of advancing to Round 2 for athletes ranked better than the cutoff increases by 18-22 percentage points after the rule change, with *p*-values between 0.042 and 0.081.

Overall, our results suggest that aligning the nominal with the effective pre-event ranks had a statistically and economically significant effect on the salience of the reference point and consequently the incentive effect of loss aversion.

**Table 5.** Difference-in-discontinuities at the elimination cutoff before and after the rule change.

| | (1) | (2) | (3) |
|---|---|---|---|
| Point estimate | 0.179 | 0.210 | 0.219 |
| | (0.102) | (0.103) | (0.125) |
| P-value | 0.081 | 0.042 | 0.080 |
| Bandwidth | [24, 37] | [24, 37] | [25, 36] |
| Effective no. of obs. (treated / controls) | 675 / 678 | 675 / 678 | 500 / 493 |
| No. of obs. | 4,790 | 4,790 | 4,167 |

Notes. The dependent variable is a dummy denoting if a ski jumper advances to Round 2 in the main event. The running variable is the pre-event rank and the cutoff is between ranks 30 and 31. The Diff-in-Disc analyses are based on local (first-order) polynomial regressions with a triangular kernel function to assign weights to the observations and common mean squared error (MSE)-optimal bandwidth selection. Point estimates report the Diff-in-Disc estimate. Standard errors are clustered at the athlete level and presented in parentheses. Conventional *p*-values are reported. The analysis in Column 2 controls for the predetermined covariates WC standing points and home event; the analysis in Column 3 additionally controls for previous event rank. All coefficients from the full regression models can be found in the supplementary replication material.

### 6.3 Treatment effects for alternative performance outcomes

Furthermore, we estimate treatment effects for alternative performance outcomes. We consider the absolute performance measures – that is, jumping distance points[22] and style points – because the probability to advance to the second round is a relative performance outcome, which might be affected by only small differences in actual performances. Moreover, an alternative explanation for our initial findings could be that differences in relative performance might be driven by performance-expectations of the judges in the panel that are raised by the pre-event ranking information, rather than by differences in the actual performances of athletes. The cutoff could serve as a reference point in judges' decision making, affecting their style point scores. In fact, several studies have documented that ski jumping judges are biased in their evaluations to some extent (e.g., Krumer et al., 2022; Zitzewitz, 2006). Therefore, we analyze treatment effects for both performance measures separately.

[22] In contrast to jumping distance in meters, jumping distance points factor in the hill size of WC events and thus allow for a cleaner comparison across competitions at different venues.

As previously, the results on jumping distance points (Table A8 in Appendix C) and style points (Table A9 in Appendix C) show no significant treatment effect for both performance measures before the rule change. However, after the rule change, we find significant effects for both measures. The results for jumping distance after the change appear in Columns 4 to 6 in Table A8 in Appendix C. Inside the largest suggested window [26, 35], athletes with pre-event ranks 26-30 achieve on average 4.3 points more than athletes with pre-event ranks 31-35. This suggests that performance differences are substantial in absolute measures as well. This also suggests that differences in performance are not predominantly driven by performance expectations of judges, since unlike style points, which is a subjective evaluation, jumping distance is an objective performance. The average style points difference between athletes with better and worse pre-event ranks is also substantial after the rule change, but the effect is only significant at the 10% level when considering the smallest windows (see Column 4 in Table A9 in Appendix C). When taking the larger windows, the size of the effect becomes slightly smaller, but significant at the 5% and 1% levels, respectively (Columns 5 and 6 in Table A9 in Appendix C). Overall, our findings suggest that the reference point is salient when the nominal pre-event ranks are identical to the effective ones, raising performance expectations among athletes and activating the incentive effect of loss aversion.

## 7 Conclusion

Nature rarely creates opportunities to observe the effect of reference-point salience on the loss aversion mechanism in effort provision. The natural experiment we have studied provides a unique opportunity to observe the behavior of highly incentivized and professional individuals in real competitive settings with an exogenous variation of reference-point salience.

We find that only when the reference point is salient, contestants with recent performances of just above the reference point perform better than those who were just below it. We attribute this result to the differences in expectations that may develop as a result of being at either side of the reference point. This finding is driven by the intuition that contestants with positive expectations (those above reference point) stand to lose more than contestants with negative expectations (those below reference point) stand to gain. However, this effect of expectation-based loss aversion is activated only when the reference point is salient.

Our findings suggest that even highly professional individuals are affected by an incremental difference between nominal and effective values that obscures an otherwise salient reference point. Our paper also emphasizes the significance of the correct framing of reference points in contests. This is because by disregarding the competitive structure of zero-sum games, one can predict that a lagging player who has more to gain should do better than the leading player who has more to lose. We offer a theoretical explanation and empirical evidence of why this prediction is not correct if players rely on performance expectations to decide on their effort provision.

Given the uniqueness of our setting, it is natural to discuss external validity of our findings. In fact, according to List (2020), this uniqueness is more of an advantage since it allows us to make the relevant test as no other settings can provide that level of relevance. Also, according to List (2020), "all results are externally valid to some settings, and no results will be externally valid to all settings" (p. 45). Nevertheless, we need to be cautious about generalizing our findings for several reasons: First, our results come from competitions among men. However, in most settings, men and women compete (cooperate) against (with) each other in mixed-gender environments. Second, ski jumping is a sport where performances are executed in a few seconds, and which requires a strong focus and specific abilities. Third, there are certainly not many other settings where the margin for mistakes is comparably small eventually causing severe injuries. Still, identifying such a significant effect among high-profile professionals suggests that expectation-based loss aversion (Kőszegi & Rabin, 2006) and the salience of a reference point (Bordalo et al., 2022) may play a significant role in human behavior in general and in highly competitive settings in particular.

Given that attention is a limited resource and therefore selective (Bordalo et al., 2022), our results suggest that strategic manipulation of reference-point salience might increase effort. For example, managers who wish that personnel exert more effort could emphasize that the exact goal is not too far away along with positive expectations of achieving it. Another possibility to affect productivity is by manipulating the perceived distance to the goal from time to time.

Finally, in the era of information explosion, where growing fractions of labor consume (limited) attentional resources (Loewenstein & Wojtowicz, 2025), we believe that salience, attention, and information will be emerging topics in economic research in the upcoming years. Thus, we anticipate an increase in theoretical contributions and controlled experiments to investigate the effect of salience on loss aversion mechanisms in different competitive environments. There are many open questions concerning the interplay between attention, reference points, and features that often occur in competitive environments, such as discouragement effects, psychological momentum, and performance under pressure, all of which are relevant for further exploration.

## Appendix A: Comparisons before and after the rule change

**Table A1.** Comparison of probabilities to advance to Round 2 for pre-event rank groups.

| | Before rule change (2015–2017 seasons) | After rule change (2018–2020 seasons) | | |
|---|---|---|---|---|
| | Mean (SD) | Mean (SD) | Difference in means | P-value |
| Pre-event rank 1–5 | 0.977 (0.149) | 0.982 (0.134) | -0.005 | 0.726 |
| Pre-event rank 6–10 | 0.946 (0.227) | 0.884 (0.321) | 0.062 | 0.014 |
| Pre-event rank 11–15 | 0.898 (0.304) | 0.848 (0.360) | 0.050 | 0.096 |
| Pre-event rank 16–20 | 0.772 (0.420) | 0.786 (0.411) | -0.014 | 0.719 |
| Pre-event rank 21–25 | 0.658 (0.475) | 0.645 (0.480) | 0.013 | 0.769 |
| Pre-event rank 26–30 | 0.540 (0.499) | 0.655 (0.477) | -0.115 | 0.010 |
| Pre-event rank 31–35 | 0.412 (0.493) | 0.468 (0.500) | -0.057 | 0.222 |
| Pre-event rank 36–40 | 0.404 (0.492) | 0.343 (0.476) | 0.062 | 0.165 |
| Pre-event rank 41–45 | 0.232 (0.423) | 0.270 (0.445) | -0.038 | 0.341 |
| Pre-event rank 46–50 | 0.195 (0.400) | 0.189 (0.392) | 0.006 | 0.873 |
| No. of obs. | 4,790 | | | |

Notes. The dependent variable is a dummy denoting if a ski jumper advances to Round 2 in the main event. Standard deviations (SD) are reported in parentheses. Reported are two-sided $p$-values of $t$-tests.

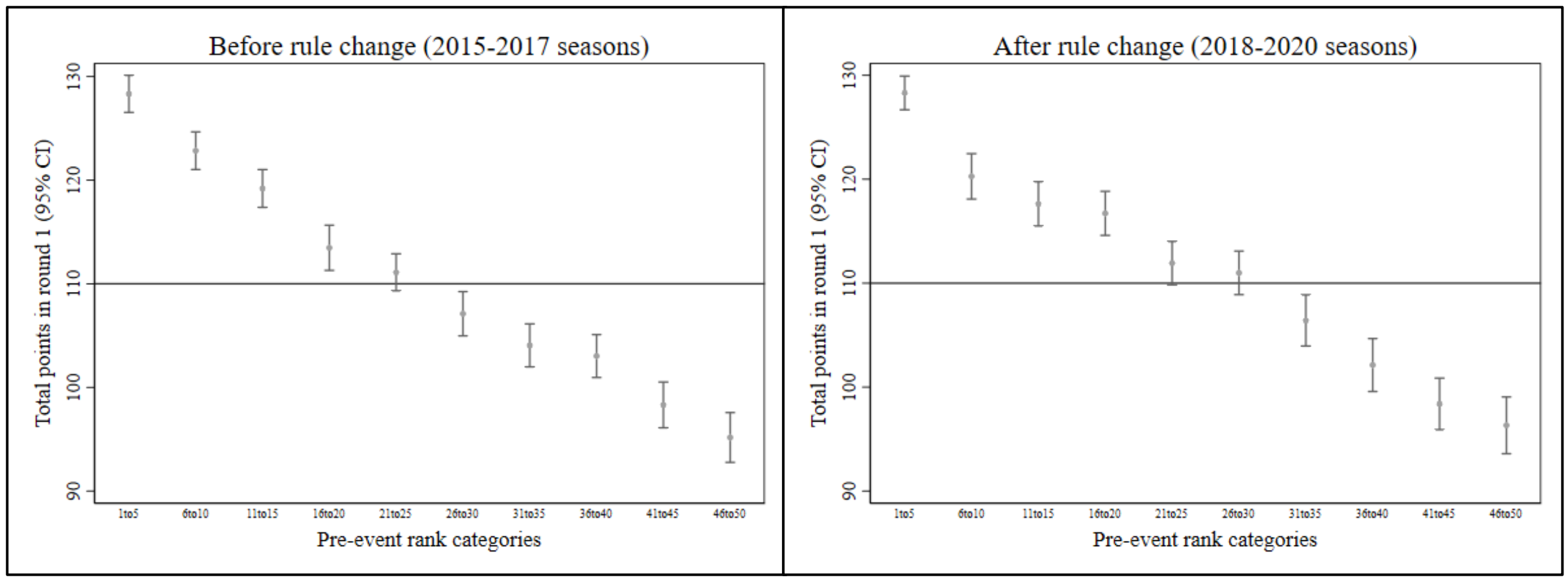

**Figure A1.** Comparison of total point scores in Round 1 as a function of ski jumpers' pre-event ranks.

**Table A2.** Comparison of total point scores in Round 1 for pre-event rank groups.

| | Before rule change (2015–2017 seasons) | After rule change (2018–2020 seasons) | | |
|---|---|---|---|---|
| | Mean (SD) | Mean (SD) | Difference in means | P-value |
| Pre-event rank 1–5 | 128.329 (14.805) | 128.303 (12.218) | 0.026 | 0.983 |
| Pre-event rank 6–10 | 122.843 (14.835) | 120.277 (16.248) | 2.566 | 0.073 |
| Pre-event rank 11–15 | 119.196 (15.043) | 117.631 (16.132) | 1.565 | 0.269 |
| Pre-event rank 16–20 | 113.473 (17.959) | 116.717 (15.571) | -3.245 | 0.039 |
| Pre-event rank 21–25 | 111.107 (14.677) | 111.929 (15.474) | -0.822 | 0.553 |
| Pre-event rank 26–30 | 107.108 (17.651) | 110.989 (15.750) | -3.881 | 0.012 |
| Pre-event rank 31–35 | 104.053 (17.037) | 106.412 (18.449) | -2.358 | 0.147 |
| Pre-event rank 36–40 | 103.026 (17.351) | 102.123 (18.869) | 0.903 | 0.585 |
| Pre-event rank 41–45 | 98.308 (18.040) | 98.380 (18.455) | -0.072 | 0.966 |
| Pre-event rank 46–50 | 95.167 (19.559) | 96.321 (20.184) | -1.154 | 0.531 |
| No. of obs. | 4,790 | | | |

Notes. The dependent variable is the total point score in Round 1 of the main event. Standard deviations (SD) are reported in parentheses. Reported are two-sided $p$-values of $t$-tests.

**Table A3.** Comparison of WC standing points for pre-event rank groups.

| | Before rule change (2015–2017 seasons) | After rule change (2018–2020 seasons) | | |
|---|---|---|---|---|
| | Mean (SD) | Mean (SD) | $z$-statistic | P-value |
| Pre-event rank 1–5 | 731.559 (451.733) | 462.597 (405.154) | 6.955 | 0.000 |
| Pre-event rank 6–10 | 417.707 (254.890) | 309.991 (294.289) | 5.195 | 0.000 |
| Pre-event rank 11–15 | 166.599 (148.355) | 231.897 (252.598) | -1.766 | 0.077 |
| Pre-event rank 16–20 | 121.498 (128.311) | 188.324 (253.588) | -1.503 | 0.133 |
| Pre-event rank 21–25 | 97.608 (112.605) | 137.435 (200.296) | -0.082 | 0.935 |
| Pre-event rank 26–30 | 86.117 (115.820) | 110.832 (176.245) | -1.392 | 0.164 |
| Pre-event rank 31–35 | 67.752 (100.954) | 95.573 (173.395) | -1.119 | 0.263 |
| Pre-event rank 36–40 | 58.901 (90.961) | 61.310 (136.083) | 2.010 | 0.044 |
| Pre-event rank 41–45 | 34.544 (74.479) | 41.274 (99.553) | -0.082 | 0.935 |
| Pre-event rank 46–50 | 34.650 (73.940) | 33.689 (125.563) | 1.671 | 0.095 |
| No. of obs. | 4,790 | | | |

Notes. The dependent variable is the ski jumpers' points in the WC standings prior to each event. Standard deviations (SD) are reported in parentheses. Since only the best 30 ski jumpers from each competition get WC points and the distribution of points is inconsistent across the finals ranks, including linear and exponential components, the variable is not normally distributed. Therefore, we use a rank-sum test to analyze differences in the distribution. Reported are $z$- and two-sided $p$-values of Mann-Whitney $U$ tests.

## Appendix B: RD validation and falsification tests

The RD approach can be further supported by validation and falsification tests, including treatment effect tests on the predetermined covariates, assessing the density of the running variable, and estimations at placebo cutoffs (Cattaneo & Titiunik, 2022; Cattaneo et al., 2023a).

As is the case in classical experiments where randomization should produce similar distributions in covariate characteristics for treated and control, one should test whether the RD treatment influences predetermined covariates. Based on the underlying assumption of the RD approach, there should be no treatment effect on predetermined covariates, and thus there should be no systematic differences between treated and control groups. Testing this also justifies our within-comparison as identification approach because it provides empirical evidence that athletes below and above the cutoff are indeed comparable. To test this, we run RD estimations on our three covariate measures, using local randomization RD with the smallest and largest window specifications as well as continuity-based RD estimations. The results, presented in Table A4, support our RD approach, showing no significant effects on these variables.

**Table A4.** RD estimates on predetermined covariates.

| | Before rule change (2015–2017 seasons) | | | |
|---|---|---|---|---|
| Variable | Window $W$ / bandwidth $h$ | Point estimate | P-value | Effective no. of obs. |
| WC standing points | $W$ [30, 31] | 10.485 | 0.694 | 51 / 53 |
| | $W$ [28, 33] | 16.919 | 0.198 | 160 / 160 |
| | $h$ [25, 36] | 6.258 | 0.869 | 320 / 316 |
| Previous event rank | $W$ [30, 31] | 2.412 | 0.360 | 39 / 47 |
| | $W$ [28, 33] | -0.144 | 0.944 | 133 / 143 |
| | $h$ [25, 36] | 1.724 | 0.455 | 276 / 273 |
| Home event | $W$ [30, 31] | 0.063 | 0.508 | 51 / 53 |
| | $W$ [28, 33] | 0.013 | 0.882 | 160 / 160 |
| | $h$ [25, 36] | 0.031 | 0.544 | 320 / 316 |
| | **After rule change (2018–2020 seasons)** | | | |
| Variable | Window $W$ / bandwidth $h$ | Point estimate | P-value | Effective no. of obs. |
| WC standing points | $W$ [30, 31] | -14.893 | 0.882 | 47 / 42 |
| | $W$ [26, 35] | 15.258 | 0.356 | 220 / 218 |
| | $h$ [24, 37] | -39.921 | 0.165 | 302 / 308 |
| Previous event rank | $W$ [30, 31] | 1.026 | 0.716 | 40 / 38 |
| | $W$ [26, 35] | -0.324 | 0.768 | 188 / 184 |
| | $h$ [22, 39] | 3.091 | 0.200 | 333 / 328 |
| Home event | $W$ [30, 31] | -0.079 | 0.350 | 47 / 42 |
| | $W$ [26, 35] | 0.026 | 0.350 | 220 / 218 |
| | $h$ [27, 34] | -0.078 | 0.224 | 174 / 171 |

Notes. Dependent variables are the predetermined covariates. The running variable is the pre-event rank and the cutoff is between rank 30 and 31. Point estimates report the difference in means of the local randomization RD analyses (with Fisherian $p$-values) or the MSE-optimal point estimates (with robust $p$-values) of local linear RD analyses. Before the rule change, the small and large window is between pre-event ranks [30, 31] and [28, 33], respectively. After the rule change, the small and large window is between pre-event ranks [30, 31] and [26, 35], respectively. The continuity-based RD analyses estimate local linear regressions with MSE-optimal bandwidth selection and with triangular kernel.

We further assess the density of the running variable by testing whether the numbers of observations on both sides of the cutoff are roughly similar (Cattaneo et al., 2023a). In principle, this should not be an issue because our running variable consists of relative ranks, which makes uneven bunching of score values unlikely. This is also a key feature of our ski jumping setting because it hinders athletes from influencing their pre-event ranks strategically. To further support this empirically, in Table A5, we present the absolute numbers of observations at the closest mass points around the cutoff. While we see some variation across pre-event ranks because athletes with the same number of points in the qualification get the same pre-event rank, we find no pattern of bunching on either side of the cutoff. We further ran binomial tests with a success probability equal to 1/2 as suggested by Cattaneo et al. (2023a), which shows no systematic difference in the number of treated and control observations.[23]

**Table A5.** Frequency distribution of mass points of the running variable around the cutoff.

| | | Before rule change (2015–2017 seasons) | After rule change (2018–2020 seasons) |
|---|---|---|---|
| | Treatment status | No. of obs. | No. of obs. |
| Pre-event rank 26 | treated | 53 | 46 |
| Pre-event rank 27 | treated | 52 | 41 |
| Pre-event rank 28 | treated | 52 | 48 |
| Pre-event rank 29 | treated | 57 | 38 |
| Pre-event rank 30 | treated | 51 | 47 |
| Pre-event rank 31 | controls | 53 | 42 |
| Pre-event rank 32 | controls | 52 | 47 |
| Pre-event rank 33 | controls | 55 | 44 |
| Pre-event rank 34 | controls | 50 | 38 |
| Pre-event rank 35 | controls | 52 | 47 |

Notes. Presented are the absolute numbers of observations at closest mass points around the cutoff, which is between pre-event rank 30 and 31.

Furthermore, we run RD estimations at placebo cutoffs below and above the actual cutoff to test for discontinuous changes in performances away from the reference point, where we would expect no treatment effect (Cattaneo & Titiunik, 2022). We choose placebo cutoffs where athletes might experience possible elimination after Round 1; that is, cutoffs between the pre-event ranks [20, 21] and [40, 41]. Moreover, the cutoff between pre-event ranks 40 and 41 would be the salient reference point for athletes before the rule change if athletes neglect the existence of the prequalified top 10 athletes and only consider their nominal ranks. Therefore, the RD analysis at cutoff [40, 41] in the sample period before the rule change also examines whether there is an effect on performances of athletes with nominal ranks that are close to the elimination cutoff as indicated in Figure 3. The RD results for the periods before and after the rule change are provided in Table A6 and A7, respectively, using the same window sizes as in our initial analyses as well as continuity-based RD analyses. We do not find discontinuous changes in advancing to Round 2 in the windows closely around the placebo cutoffs (Columns 1, 2, 5, and 6 in Tables A6 and A7) or when using local linear RD (Columns 4 and 8 in Tables A6 and A7). Only when considering the largest windows in local randomization RD, we find significant differences for $c = 40.5$ before the rule change (Column 7 in Table A6) and $c = 20.5$ after the rule change (Column 3 in Table A7). However, in both cases, the treated and untreated observations also differ regarding their predetermined covariates. Overall, the findings support the assumption that the probability of advancing to Round 2 as a function of pre-event ranks is continuous at ranking positions where the real treatment is assumed to be absent.

[23] The binomial tests report *p*-values of 0.922 and 1.000 for the smallest and largest windows before the rule change, respectively; and *p*-values of 0.672 and 0.962 for the smallest and largest windows after the rule change, respectively.

**Table A6.** RD estimates on advancing to Round 2 at placebo cutoffs.

| | Before rule change (2015–2017 seasons) | | | | | | | |
|---|---|---|---|---|---|---|---|---|
| | (1) | (2) | (3) | (4) | (5) | (6) | (7) | (8) |
| Point estimate | 0.050 | 0.047 | 0.082 | 0.020<br>(0.073) | 0.119 | 0.071 | 0.135 | 0.070<br>(0.117) |
| P-value | 0.776 | 0.544 | 0.134 | 0.701 | 0.310 | 0.290 | 0.016 | 0.532 |
| Window / bandwidth | [20, 21] | [19, 22] | [18, 23] | [14, 27] | [40, 41] | [39, 42] | [38, 43]* | [38, 43] |
| Placebo cutoff | 20.5 | 20.5 | 20.5 | 20.5 | 40.5 | 40.5 | 40.5 | 40.5 |
| Effective no. of obs. (treated / controls) | 53/51 | 106/103 | 158/155 | 371/368 | 58/50 | 112/98 | 164/152 | 164/152 |

Notes. The dependent variable is a dummy denoting if a ski jumper advances to Round 2 in the main event. The running variable is the pre-event rank. Columns 1–3 and 5–7 report local randomization RD analyses and the difference in means as point estimates with Fisherian $p$-values. Columns 4 and 8 present local linear regressions with MSE-optimal bandwidth selection and with triangular kernel. Reported are the MSE-optimal point estimates with standard errors clustered at the athlete level presented in parentheses and robust $p$-values. *This window does not pass covariate balance tests.

**Table A7.** RD estimates on advancing to Round 2 at placebo cutoffs.

| | After rule change (2015-2017 seasons) | | | | | | | |
|---|---|---|---|---|---|---|---|---|
| | (1) | (2) | (3) | (4) | (5) | (6) | (7) | (8) |
| Point estimate | 0.136 | 0.077 | 0.141 | 0.123<br>(0.075) | -0.091 | -0.067 | 0.073 | -0.164<br>(0.146) |
| P-value | 0.282 | 0.340 | 0.002 | 0.142 | 0.482 | 0.410 | 0.110 | 0.327 |
| Window / bandwidth | [20, 21] | [19, 22] | [16, 25]* | [14, 27] | [40, 41] | [39, 42] | [36, 45] | [38, 43] |
| Placebo cutoff | 20.5 | 20.5 | 20.5 | 20.5 | 40.5 | 40.5 | 40.5 | 40.5 |
| Effective no. of obs. (treated / controls) | 42/46 | 84/92 | 210/214 | 298/301 | 44/44 | 84/85 | 213/215 | 123/129 |

Notes. The dependent variable is a dummy denoting if a ski jumper advances to Round 2 in the main event. The running variable is the pre-event rank. Columns 1–3 and 5–7 report local randomization RD analyses and the difference in means as point estimates with Fisherian *p*-values. Columns 4 and 8 present local linear regressions with MSE-optimal bandwidth selection and with triangular kernel. Reported are the MSE-optimal point estimates with standard errors clustered at the athlete level presented in parentheses and robust *p*-values. *This window does not pass covariate balance tests.

# Appendix C: Treatment effects on alternative performance outcomes

**Table A8.** Local randomization RD estimates on jumping distance points.

| | Before rule change (2015–2017 seasons) | | | After rule change (2018–2020 seasons) | | |
|---|---|---|---|---|---|---|
| | (1) | (2) | (3) | (4) | (5) | (6) |
| Point estimate | 1.235 | 0.597 | -0.247 | 7.828 | 5.603 | 4.293 |
| P-value | 0.562 | 0.744 | 0.862 | 0.010 | 0.014 | 0.002 |
| Window | [30, 31] | [29, 32] | [28, 33] | [30, 31] | [29, 32] | [26, 35] |
| Effective no. of obs. (treated / controls) | 51 / 53 | 108 / 105 | 160 / 160 | 47 / 42 | 85 / 89 | 220 / 218 |
| No. of obs. | | 2,629 | | | 2,161 | |

Notes. The dependent variable is the jumping distance in points in Round 1 of the main event. The running variable is the pre-event rank and the cutoff is between rank 30 and 31. The windows for RD analyses in columns 3 and 6 derive from optimal window selection based on the predetermined covariates WC standing points, previous event rank, and home event. Point estimates report the difference in means and *p*-values derived from Fisherian simulation-based methods.

**Table A9.** Local randomization RD estimates on style points.

| | Before rule change (2015–2017 seasons) | | | After rule change (2018–2020 seasons) | | |
|---|---|---|---|---|---|---|
| | (1) | (2) | (3) | (4) | (5) | (6) |
| Point estimate | 0.094 | -0.004 | -0.031 | 0.921 | 0.856 | 0.720 |
| P-value | 0.812 | 0.958 | 0.888 | 0.096 | 0.020 | 0.002 |
| Window | [30, 31] | [29, 32] | [28, 33] | [30, 31] | [29, 32] | [26, 35] |
| Effective no. of obs. (treated / controls) | 51 / 53 | 108 / 105 | 160 / 160 | 47 / 42 | 85 / 89 | 220 / 218 |
| No. of obs. | | 2,629 | | | 2,161 | |

Notes. The dependent variable is the style points in Round 1 of the main event. The running variable is the pre-event rank and the cutoff is between ranks 30 and 31. The windows for RD analyses in Columns 3 and 6 derive from optimal window selection based on the predetermined covariates WC standing points, previous event rank, and home event. Point estimates report the difference in means and *p*-values derived from Fisherian simulation-based methods.